\documentclass[12pt]{iopart}
\usepackage{amssymb,bm}
\usepackage{graphicx}
\usepackage{amsfonts}
\usepackage{amssymb}
\usepackage{bbm}
\usepackage{enumitem}

\usepackage{graphicx}
\usepackage{subfigure}
\usepackage{color}

\newcommand{\bra}[1]{\langle #1 |}
\newcommand{\ket}[1]{| #1 \rangle}

\newcommand{\eqref}[1]{(\ref{#1})}

\usepackage{subfigure}
\usepackage{color}

\usepackage{graphicx}
\usepackage{bbm}
\usepackage{enumitem}

\usepackage[usenames,svgnames]{xcolor}
\usepackage{url}
\usepackage{umoline}
\usepackage[breaklinks]{hyperref}
\hypersetup{
     colorlinks=true,       		% false: boxed links; true: colored links
     linkcolor=Salmon,          	% color of internal links
     citecolor=blue,            % color of links to bibliography
     filecolor=blue,      		% color of file links
     urlcolor=cyan,           	% color of external links
 }

\newtheorem{theorem}{Theorem}[section]

\newcommand{\Oorderof}{\mathcal{O}}
\newcommand{\orderof}[1]{\Oorderof(#1)}

\newcommand{\co}[0]{ {\rm c}}
\newcommand{\for}[0]{\quad {\rm for} \quad}

\newcommand{\Prod}[0]{ {\rm Prod}}
\newcommand{\rank}[0]{ {\rm rank}}

\begin{document}

\title[  ]{Exponential bound on information spreading induced by quantum many-body
dynamics with long-range interactions}

\author{Tomotaka Kuwahara}

\address{WPI, Advanced Institute for Materials Research, Tohoku University, Sendai 980-8577, Japan}
\ead{tomotaka.kuwahara.e8@tohoku.ac.jp}

\begin{abstract}

The dynamics of quantum systems strongly depends on the local structure of the Hamiltonian.
For short-range interacting systems, the well-known Lieb-Robinson bound defines the effective light cone 
with an exponentially small error with respect to the spatial distance, 
whereas we can obtain only polynomially small error for distance in long-range interacting systems.
In this paper, we derive a qualitatively new bound for quantum dynamics by considering how many spins can correlate with each other after time evolution.
Our bound characterizes \textit{the number of spins} which support the many-body entanglement with exponentially small error and is valid for large class of Hamiltonians including long-range interacting systems.     
To demonstrate the advantage of our approach in quantum many-body systems, we apply our bound to prove several fundamental properties which have not be derived from the Lieb-Robinson bound.

\end{abstract}

%cover letter 
%


\maketitle

%\section{Introduction}\label{Intro}
\section{Introduction}

The fundamental features of quantum many-body systems are strongly restricted by the local nature of Hamiltonian.
Such restrictions give us a lot of useful information in analyzing universal properties of matters.
One of the prominent examples is the Lieb-Robinson bound~\cite{ref:LR-bound72,PhysRevLett.97.050401},
which characterizes the velocity of information propagation in non-relativistic quantum systems; 
in other words, we can define an approximate ``light cone'' with an exponentially small error.
Based on the Lieb-Robinson bound, we can grasp fundamental restrictions to quantum dynamics:
entropy production rate after quench~\cite{
PhysRevLett.111.170501,
PhysRevA.76.052319,
PhysRevLett.97.150404,
PhysRevA.78.010306}, 
%%%%%%%%%%%%%%%%%%%%%%%%%%%%%%%%%%%%%%%%%%%%%%%%%%%%%%%%%%%%%%%%%%%%%%%%%%%%%%
entanglement growth~\cite{
PhysRevLett.99.167201,
PhysRevLett.102.017204,
PhysRevX.3.031015,
PhysRevLett.113.187203,
lauchli2008spreading,
PhysRevA.80.052319,
PhysRevA.81.062107,
PhysRevA.84.032309},
%%%%%%%%%%%%%%%%%%%%%%%%%%%%%%%%%%%%%%%%%%%%%%%%%%%%%%%%%%%%%%%%%%%%%%%%%%%%%%
complexity of quantum simulation~\cite{
PhysRevB.73.094423,
ref:Osborne2006-Efficient,
ref:Osborne2007-Adiabatic,
ref:Hastings2009-Adiabatic,
PhysRevLett.101.070503}, and so on.
Moreover, the Lieb-Robinson bound also provides us powerful analytical tools to give foundations of quantum many-body systems,
from condensed matter physics to statistical mechanics: 
%%%%%%%%%%%%%%%%%%%%%%%%%%%%%%%%%%%%%%%%%%%%%%%%%%%%%%%%%%%%%%%%%%%%%%%%%%%%%%
Lieb-Schultz-Mattis theorem~\cite{
ref:LSM-Hastings04,
ref:Nachtergaele2007-LSM},
%%%%%%%%%%%%%%%%%%%%%%%%%%%%%%%%%%%%%%%%%%%%%%%%%%%%%%%%%%%%%%%%%%%%%%%%%%%%%%
exponential decay of bi-partite correlation~\cite{
ref:Hastings2004-Markov,
ref:Hastings2006-ExpDec,
ref:Nachtergaele2006-LR}, 
%%%%%%%%%%%%%%%%%%%%%%%%%%%%%%%%%%%%%%%%%%%%%%%%%%%%%%%%%%%%%%%%%%%%%%%%%%%%%%
stability of topological order to perturbation~\cite{
bravyi2010topological,
bravyi2011short,
PhysRevLett.97.050401,
PhysRevB.72.045141,
michalakis2013stability},
%%%%%%%%%%%%%%%%%%%%%%%%%%%%%%%%%%%%%%%%%%%%%%%%%%%%%%%%%%%%%%%%%%%%%%%%%%%%%%
quantization of the Hall conductance~\cite{hastings2015quantization,hastings2010quasi}, 
%%%%%%%%%%%%%%%%%%%%%%%%%%%%%%%%%%%%%%%%%%%%%%%%%%%%%%%%%%%%%%%%%%%%%%%%%%%%%%
thermalization problem~\cite{
ref:Eisert2015-Thermal,
ref:Mueller2013-Thermal}, 
%%%%%%%%%%%%%%%%%%%%%%%%%%%%%%%%%%%%%%%%%%%%%%%%%%%%%%%%%%%%%%%%%%%%%%%%%%%%%%
equivalence of the statistical ensembles~\cite{ref:Brandao2015-Berry-Esseen}, etc. 
In these results, the locality of interactions plays essential roles. 
Thus,  the principle of locality has shed new light on our understanding of fundamental many-body physics.

%Locality analysis

More recently, with the progress of experimental technology~\cite{cheneau2012light,langen2013local,richerme2014non,jurcevic2014quasiparticle}, 
there has been considerable interest in the potential of the locality analysis in long-range interacting systems, both from theoretical~\cite{ref:Arad-kuwahara,kuwahara2015floquet, kuwahara2015local, PhysRevX.3.031015,PhysRevLett.113.030602,PhysRevLett.114.157201,PhysRevLett.112.210601,PhysRevLett.111.207202,PhysRevB.90.174204,PhysRevLett.111.260401,PhysRevLett.116.120401} and experimental~\cite{richerme2014non,jurcevic2014quasiparticle} 
viewpoints. 
In such systems, we can also define the approximate light cone as in the case of the short-range interacting systems.
However, the light-cone is usually nonlinear to the time except some special cases~\cite{PhysRevLett.114.157201}, and moreover the transport of information can be bounded only polynomially~\cite{PhysRevLett.113.030602, PhysRevLett.111.260401,ref:Hastings2006-ExpDec,richerme2014non,PhysRevLett.114.157201} with respect to the spatial distance outside the light cone. 
% this bound is much looser than the exponentially strong Lieb-Robinson bound in short-range interacting systems.
The primary reason is that the Lieb-Robinson bound focuses on the velocity of the information transfer, whereas the long-range interacting systems can transport information immediately in principle.
In this way, the causality allows us to analyze the system in the looser way in comparison with the case of the short-range interacting systems.
This indicates that we may not grasp all the restrictions due to the locality of the Hamiltonian in terms of only the spatial distance. 
Here, we use the term of ``locality" in more broader meanings~\cite{kuwahara2015local} as distinguished from the spatial locality, i.e., \textit{
to what extent can quantum systems be described by a collection of local
degrees of freedom, which are only loosely correlated
with each other?}

%
%the causality of the system itself give us a restricted applications to various fundamental properties in compared with the case of the short-range interacting systems.

\begin{figure}[tt]
\centering
\includegraphics[clip, scale=0.5]{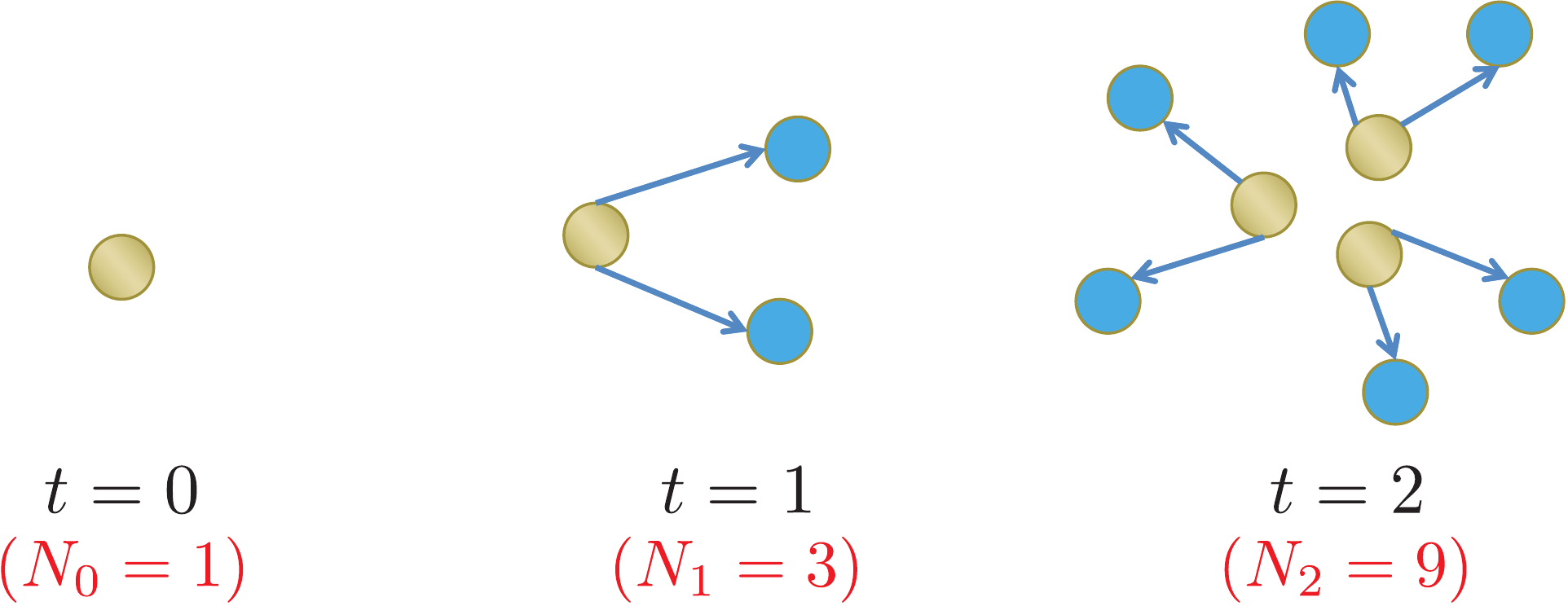}
\caption{\footnotesize  Schematic picture of classical information spreading. 
We consider a classical system with $N$ elements and assume that each of the elements 
is \textit{connected to arbitrary ones}. 
 We start from one information source (e.g., coded by $0$ or $1$) and consider the process that each element which has received the information can send the information to other $2$ elements per unit of time.
For example, after a unit of time, $2$ elements receive the information and then totally $3$ elements becomes the next information source. In the same way,  after two units of time, totally $9$ elements share the information.
Repeatedly,  the number of receivers　 after $n$ units of time, say $N_n$, increases as $3^n$. 
By analogy, we also expect that the information spreading of spins should be similarly suppressed in the quantum many-body dynamics.
}
\label{fig:Fig2_chap6}
\end{figure}

In the present paper, we give a qualitatively new bound for dynamical properties in terms of \textit{the number of spins instead of the spin-spin distance};
roughly speaking, we focus on how many spins can correlate with each other after time evolution. 
In order to make our concept clear, we first consider a classical process in which a source transfers information to receivers.
We now assume that each of the elements sends information to \textit{any} other 2 elements per unit of time (see Fig.~\ref{fig:Fig2_chap6}). 
Then, the number of elements which share the information can be bounded from above by $e^{\orderof{t}}$. 
Our main purpose is to give a quantum version of the bound in the form of an operator inequality. 
As we shall see shortly, our new bound characterizes the number of particles which support many-body entanglement such as the topological order~\cite{PhysRevLett.97.050401} and the macroscopic entanglement~\cite{shimizu2005detection,frowis2012}; this indicates that 
the global entanglement induced by a small-time evolution should be suppressed up to exponentially small error.
It allows us to obtain novel strong properties on various kinds of fundamental physics in quantum many-body systems which have not  be derived from the Lieb-Robinson bound.

%%%%%%%%%%%%%%%%%%%%%%%%%%%%%%%%%%%%%%%%%%%%%%%%%%%%%%%%%%%%%%%%%%%%%%%%%%%%%%%%%%%%%%%%%%%%%%%%%%%%%%%%%%%%%%%%%%%%%%%%%%%%%%%%%%%%%%%%%%%%%%%%%%%%%%%%%%%%%%%%%%%%%%%%%%%%%%%%%%%%%%%%%%%%%%%%%%%%%%%%%%%%%%%%%%%%%%%%%%%%%%%%%%%%%%%%%%%%%%%%%%%%%%%%%%%%%%%%%%%%%%%%%%%%%%%%%%%%%%%%%%%%%%%%%%%%%%%%%%%%%%%%%%%%%%%%%%%%%%%%%%%%%%%%%%%%%%%%%%%%%%%%%%%%%%%%%%%%%%%%%%%%%%%%%%%%%%%%%%%%%%%%%%%%%%%%%%%%%%%%%%%%%%%%%%%%%%%%%%%%%%%%%%%%%%%%%%%%%%%%%%%%%%%%%%%%%%%%%%%%%%%%%%%%%%%%%%%%%%%%%%%%%%%%%%%%%%%%%%%%%%%%%%%%%%%%%%%%%%%%%%%%%%%%%%%%%%%%%%%%%%%%%%%%%%%%%%%%%%%%%%%%%%%%%%%%%%%%%%%%%%%%%%%%%%%%%%%%%%%%%%%%%%%%%%%%%%%%%%%%%%%%%%%%%%%%%%%%%%%%%%%%%%%%%%%%%%%%%%%%%%%%%%%%%%%%%%%%%%%%%%%%%%%%%%%%%%%%%%%%%%%%%%%%%%%%%%%%%%%%%%%%%

\section{Model and formalism}\label{sec:set_up}

We consider a spin system of finite volume with each spin having a $d$-dimensional Hilbert space and label each spin by $i=1,2,\ldots  N$.　
We denote the set of all spins by $\Lambda:=\{1, 2, \ldots , N\}$
We denote partial sets of sites by $X$, $Y$, $Z$ and so on and the cardinality of $X$, that is, the number of sites contained in $X$, by $|X|$
 (e.g. $X=\{i_1,i_2,\ldots, i_{|X|}\}$).
%We also denote the complementary subsets of $X$, $Y$ and $Z$ by $X^\co$, $Y^\co$ and $Z^\co$, respectively; in other words, $X\oplus X^\co$ comprises the total system.
We here define the \textit{$q$-local operator} $\Gamma^{(q)}$ as follows:
\begin{eqnarray}
\Gamma^{(q)}=\sum_{|X|\le q }\gamma_X \quad \textrm{($q$-locality)}, \label{k_locality_def}
\end{eqnarray}
where each of the $\{\gamma_X\}$ is supported in a finite set $X \subset \Lambda$.  
In other words, the $q$-local operator contains up to $q$-body coupling. 

Here, we assume systems which are governed by $k$-local Hamiltonians with $k=\orderof{1}$: 
\begin{eqnarray}
H =\sum_{|X|\le k }h_X .
\end{eqnarray}
We assume the time-independence of the Hamiltonian for the simplicity, but the discussion can be generalized to the time-dependent Hamiltonian $H(t)$. 
Note that we make no assumption on the geometry of the system, and the coupling can be arbitrarily long ranged. 
Instead, as a normalization factor, we introduce the parameter $g$ of the Hamiltonian:
 \begin{eqnarray}
\sum_{X: X\ni i} \|h_X\| \le g  \quad {\rm for }\quad  \forall i\in \Lambda \quad \textrm{($g$-extensiveness)}  \label{g-extensive}
\end{eqnarray}
with $\|\cdots\|$ the operator norm (i.e., the maximum singular value of the operator); we refer to that the Hamiltonian is $g$-extensive if it satisfies the condition \eqref{g-extensive}.
This implies that the energy associated with one spin is bounded by a finite value $g$.
Note that the norm of the Hamiltonian $\|H\|$
increases at most linearly with the system size $N$, namely 
  \begin{eqnarray}
\|H\|=\left\| \sum_{X}  h_X \right\| \le \sum_{X} \| h_X\| \le \sum_{i=1}^N \sum_{X: X\ni i} \|h_X\| \le  \sum_{i=1}^N g =gN.
\end{eqnarray}
We notice that the class of $k$-local Hamiltonians covers almost all realistic quantum many-body systems
not only with short-range interactions but also with long-range interactions.

For the basic analysis of the $k$-local Hamiltonian, we utilize the following theorem:
\begin{theorem}\label{k_local_fund}
Let $H$ be a $k$-local $g$-extensive Hamiltonian and $\Gamma^{(q)}$ be a $q$-local operator; note that the $\Gamma^{(q)}$ may not be extensive as in \eqref{g-extensive}.
Then, for an arbitrary positive integer $q$, we can obtain 
\begin{eqnarray}
\| [H,  \Gamma^{(q)}] \| \le\lambda  \frac{q}{k}\|\Gamma^{(q)}\| \quad {\rm with } \quad \lambda := 6gk^2.
\label{f:ineq}
\end{eqnarray}
\end{theorem}
Note that the operator $[H,  \Gamma^{(q)}]$ is still at most $(k+q)$-local. 
In the case where $\Gamma^{(q)}$ is supported in a local subset $Z\subset \Lambda$, namely $\Gamma^{(q)}=\gamma_Z$ ($|Z|=q$), we can simply prove the theorem as follows: 
 \begin{eqnarray}
 \label{eq:gamma_z}
\fl \| [H,  \gamma_Z] \| \le \sum_{X: X\cap Z \neq \emptyset} \|[h_X, \gamma_Z]\| \le  \sum_{i\in Z}\sum_{X:X\ni i} 2\|h_X\| \cdot \| \gamma_Z\| \le 2g |Z|\cdot \|\gamma_Z\|\le 6gq k\|\gamma_Z\|.
\end{eqnarray}
When $\Gamma^{(q)}$ is a general $q$-local operator, however, the proof cannot be given in a simple way but a bit technical.
To show the point, we expand a $q$-local operator $\Gamma^{(q)}$ as 
 \begin{eqnarray}
\Gamma^{(q)} = \sum_{|Z|\le q} \gamma_Z   \nonumber.
\end{eqnarray}
Now, the difficulty lies in the fact that we cannot utilize the following simple estimation; we have
  \begin{eqnarray}
\| [H,\Gamma^{(q)}]\| \le  \sum_{|Z|\le q}\| [H, \gamma_Z ] \| \le 2gq \sum_{|Z|\le q}\|\gamma_Z\|  , \nonumber
\end{eqnarray}
whereas we cannot generally ensure
  \begin{eqnarray}
\sum_{|Z|\le q}\|\gamma_Z\| \propto \|\Gamma^{(q)}\| .\nonumber
\end{eqnarray}
For example, let us consider the following $2$-local operator 
 \begin{eqnarray}
\Gamma^{(2)}=\frac{1}{N}\sum_{i<j} \gamma_{i,j} \sigma_i^z \otimes \sigma_j^z,   \nonumber
\end{eqnarray}
where $\{\gamma_{i,j}\}_{i,j=1}^N$ are  uniform random numbers from $-1$ to $1$.
For this operator, we can obtain $\sum_{i<j}\| \gamma_{i,j} \sigma_i^z \otimes \sigma_j^z\| /N =\orderof{N}$, but $\|\Gamma^{(2)}\|=\orderof{\sqrt{N}}$, and hence
 \begin{eqnarray}
 \| [H,\Gamma^{(2)}]\| &\le  \sum_{|Z|\le 2}\| [H, \gamma_Z ] \| \le 4g\sum_{i<j} \frac{\| \gamma_{i,j} \sigma_i^z \otimes \sigma_j^z\| }{N} =4g  \orderof{\sqrt{N}} \|\Gamma^{(2)}\| ,\nonumber
\end{eqnarray}
where the second inequality comes from \eqref{eq:gamma_z}.
This is much looser than the inequality~\eqref{f:ineq} which comes from Theorem~\ref{k_local_fund}.
We give the full proof in \ref{Appendix_A}.

%{~}

\section{Main results}
In order to mathematically apply the classical discussion on the information sharing to quantum cases, we consider the dynamics of the $k$-locality of operators. 
We initially consider a $q_0$-local operator $\Gamma^{(q_0)}$ and investigate its time evolution:
$
\Gamma^{(q_0)}(t) = e^{-iHt} \Gamma^{(q_0)}  e^{iHt}.
$
After a time evolution, the operator $\Gamma^{(q_0)}(t)$ will be no longer a $q_0$-local operator but may be approximated by another $q$-local operator with $q\ge q_0$, say $\Gamma_t^{(q)}$.
We now regard $q_0$ and $q$ as the numbers of particles which share the information at the times $0$ and $t$ as in Fig.~\ref{fig:Fig2_chap6}, respectively.
We then expect that the approximation can be rapidly improved beyond $q \gtrsim q_0 e^{\orderof{t_0}}$ from the classical discussion.
Indeed, we prove the following theorem for the minimal error of the approximation:

\begin{theorem}\label{Thm:exp_dynamics}
Let $\mathcal{U}(q)$ be a set of $q$-local operators and consider an arbitrary $q_0$-local operator $\Gamma^{(q_0)}\in\mathcal{U}(q_0) $.
Then, for an arbitrary real-time evolution with $t>0$, there exists a $q$-local operator $\Gamma_t^{(q)}$ which approximates the operator $\Gamma^{(q_0)}(t)$ with the following error:
\begin{eqnarray}
\inf_{\Gamma_t^{(q)} \in \mathcal{U}(q)}\bigl( \|\Gamma^{(q_0)}(t)  -\Gamma_t^{(q)}   \| \bigr)  
\le 8 \|\Gamma^{(q_0)}\|  \lceil \kappa t \rceil 
\exp \biggl [ -\frac{1}{\xi} \Bigl ( \frac{q}{r_t} -q_0 \Bigr) \biggr] \label{bound_chap6_second}
\end{eqnarray}
with $\kappa=4\lambda$, $\xi=k/\log2$ and $r_t=2^{\lceil \kappa t \rceil}-1$, where $\lceil \cdots \rceil $ denotes the ceiling function. 
The same inequality holds for $t<0$ by replacing $t$ with $|t|$.
\end{theorem}

Because the function $r_t$ increases as $e^{\orderof{t}}$, the time-evolution of $\Gamma^{(q_0)}$ can be well approximated by  a $(q_0e^{\orderof{t}})$-local operator.  
The upper bound is meaningful as long as $t\lesssim \log N$, which 
 is qualitatively consistent with the threshold time of the breakdown of the Lieb-Robinson bound for long-range interacting Hamiltonians. 
As in the case of the Lieb-Robinson bound~\cite{PhysRevLett.113.030602,PhysRevLett.114.157201}, we might improve the present theorem by 
explicitly considering a spatial structure of the system, for example, the power-law decay of interaction.

We also mention the case where the system is governed by a short-range interacting Hamiltonian.
In this case, we can obtain much stronger restrictions~\cite{PhysRevLett.97.050401}.
Let us consider an operator $\Gamma_L$ which is supported in a region $L$. 
After a short time, the operator $\Gamma_L(t)$ is no longer supported in the region $L$, but is approximately supported in some region having distance $l$ from $L$. 
The Lieb-Robinson bound ensure that the accuracy of this approximation becomes precise exponentially as the distance $l$ increases beyond $\orderof{t}$. In other words, the support of $\Gamma_L(t)$ enlarges at most as ${\rm Poly} (t)$ instead of $e^{\orderof{t}}$.

%Note that we do not place any assumptions on the operator $\Gamma^{(q_0)}$.

{~}\\
\textit{Proof of Theorem~\ref{Thm:exp_dynamics}.}
For the proof,
we first obtain the upper bound for a small-time evolution: 
 \begin{eqnarray}
\inf_{\Gamma_t^{(q)} \in \mathcal{U}(q)}\bigl( \| \Gamma^{(q_0)}(t) -\Gamma_t^{(q)} \| \bigr) \le 2^{q_0/k}\cdot  \frac{(\kappa t/2)^{(q-q_0)/k}}{1-\kappa t/2} \|\Gamma^{(q_0)}\|  \label{bo1}
\end{eqnarray}
for $t< 2/\kappa$.  
In the derivation of \eqref{bo1}, we use the direct expansion of $e^{-i H t} \Gamma^{(q_0)} e^{i H t}$ according to the Hadamard lemma of the form
  \begin{eqnarray}
\Gamma^{(q_0)}(t)  = \sum_{m=0}^\infty \frac{(-i t)^m}{m!} \overbrace{ [H,[H, \cdots [H}^m ,\Gamma^{(q_0)}]]\cdots ].  \label{Hadmal_expansion_expd2}
\end{eqnarray}
We terminate the expansion~\eqref{Hadmal_expansion_expd2} at $m=m_0$ so that the expanded operator may be $q$-local.
We then estimate the error due to the termination to prove the bound \eqref{bo1}.
We show the full proof in \ref{Bound for a small-time evolution}.

\begin{figure}[ttt]
\centering
\includegraphics[clip, scale=0.44]{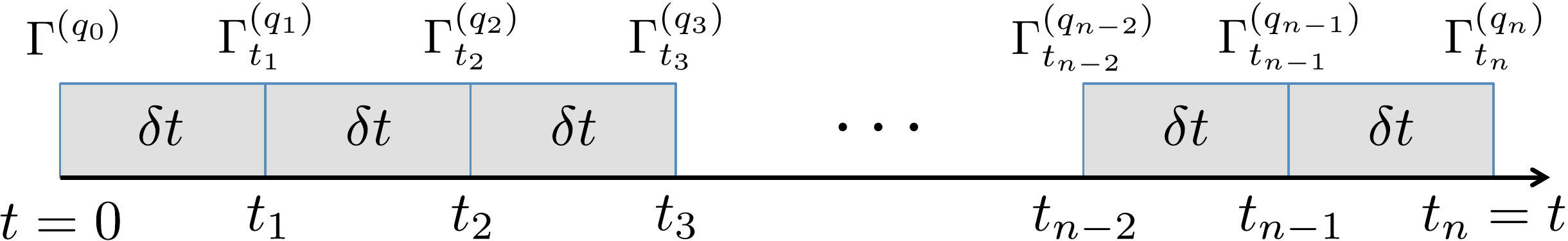}
\caption{\footnotesize  Schematic picture of the proof of Theorem~\ref{Thm:exp_dynamics}. 
For small-time evolutions, we can obtain the approximation of $\Gamma^{(q_0)}(t)$ directly by the use of the time expansion as in Eq.~\eqref{Hadmal_expansion_expd2}.
For long-time evolutions, we split the total time range $[0,t]$ into $n$ intervals with a length $\delta t$.
In each of the intervals, we can apply the result for small-time evolutions, and hence we can connect the approximations as follows; 
we first approximate $\Gamma^{(q_0)}(\delta t)$ by $\Gamma^{(q_1)}_{t_1}$, second approximate $\Gamma^{(q_1)}_{t_1}(\delta t)$ by $\Gamma^{(q_2)}_{t_2}$ and so on. Then, we can finally prove that the operator $\Gamma^{(q_n)}_{t_n}$ satisfies the inequality~\eqref{bound_chap6_second} by choosing the set $\{q_m,\Gamma^{(q_m)}_{t_m}\}_{m=1}^n$ appropriately.
}
\label{fig:Fig1_chap6}
\end{figure}

We, however, cannot utilize the expansion~\eqref{Hadmal_expansion_expd2} in order to obtain a meaningful bound for $t>2/\kappa$.
In obtaining the inequality~\eqref{bound_chap6_second}, we will have to utilize the fact that $e^{-iH t}$ is unitary~\footnote{Note that the bound~\eqref{bo1} can be also applied to the imaginary time evolution $e^{H t} \Gamma^{(q_0)} e^{-H t}$ without the unitarity condition. The approximation by the finite expansion of $e^{H t} \Gamma^{(q_0)} e^{-H t}$ will becomes less accurate beyond a certain time $t_c$, which comes from the fact that the norm of $e^{H t} \Gamma^{(q_0)} e^{-H t}$ rapidly increases for $t\ge t_c$. 
Without the unitarity, we cannot arrive at the inequality~\eqref{bound_chap6_second} from \eqref{bo1}.}.
 For this purpose, we split the time range $[0,t]$ into $n$ intervals (Fig.~\ref{fig:Fig1_chap6}) such that 
$t/n \le 1/\kappa$.
We here denote the length of the interval $t/n$ by $\delta t$:
 \begin{eqnarray}
\delta t := \frac{t}{n} \le \frac{1}{\kappa}\quad {\rm with } \quad n:= \lceil \kappa t \rceil  . \label{Definition_delta_t_chap6}
\end{eqnarray}
We also define $t_m := m\delta t$ for $m=0,1,2,\ldots n$ with $t_n=t$. 
Note that in each interval we can now apply the upper bound~\eqref{bo1}.

In the following, we connect the approximations of $\Gamma^{(q_0)}(t)$ from the first interval $[0,t_1)$ to the last interval $[t_{n-1},t_n]$. 
We first approximate the time evolution $\Gamma^{(q_0)}(\delta t)$ with a $q_1$-local operator $\Gamma^{(q_1)}_{t_1}$.
We second approximate $\Gamma^{(q_1)}_{t_1}(\delta t)$ with a $q_2$-local operator $\Gamma^{(q_2)}_{t_2}$.
By sequentially repeating this process, we define a set of operators $\{\Gamma^{(q_m)}_{t_m}\}_{m=1}^n$ so that they may approximately satisfy
$
\| \Gamma^{(q_m)}_{t_m}(\delta t) - \Gamma^{(q_{m+1})}_{t_{m+1}}\|\simeq 0,
$
respectively, where $q_n\le q$.
We thus obtain the approximation of $\Gamma^{(q_0)}(t)$ by the use of the set $\{\Gamma^{(q_m)}_{t_m}\}_{m=1}^n$:
\begin{eqnarray}
&\|\Gamma^{(q_n)}_{t_n} -\Gamma^{(q_0)}(t) \| \nonumber\\
 =&\biggl \| \sum_{m=1}^{n}  \bigl[\Gamma^{(q_m)}_{t_m} (t_n-t_m) - \Gamma^{(q_{m-1})}_{t_{m-1}} (t_n-t_m+\delta t)  \bigr] \biggr\| \nonumber\\
\le&  \sum_{m=1}^{n}  \bigl \|\Gamma^{(q_m)}_{t_m} (t_n-t_m)   -  \Gamma^{(q_{m-1})}_{t_{m-1}}(t_n-t_m+\delta t)  \bigr\| \nonumber\\
=& \sum_{m=1}^{n}  \bigl \| \Gamma^{(q_m)}_{t_m} -  \Gamma^{(q_{m-1})}_{t_{m-1}} (\delta t)   \bigr\|  , \label{Gamma_l_m_t_m_Delta2}
\end{eqnarray}
 where we used the equality $t_{m+1}-t_m = \delta t$ in the first line, the norm invariance during the time evolution in the third line, and we set $ \Gamma^{(q_{0})}_{t_{0}} =: \Gamma^{(q_0)}$.
Note that we use the unitarity of the time evolution in the third equality.

We can then show the appropriate choice of the set $\{q_m, \Gamma^{(q_m)}_{t_m}\}_{m=1}^n$.
From the inequality~\eqref{bo1}, we can prove that there exists a set $\{q_m, \Gamma^{(q_m)}_{t_m}\}_{m=1}^n$ such that: 
\begin{eqnarray}
\bigl \| \Gamma^{(q_m)}_{t_m} -  \Gamma^{(q_{m-1})}_{t_{m-1}} (\delta t)   \bigr\|  \le \Delta  (\Delta +1)^{m-1} \|\Gamma^{(q_0)}\|   \label{G02}
\end{eqnarray}
for $m=1,2,\ldots,n$, respectively, with 
$
\Delta := 4 \exp \bigl [ -\frac{1}{\xi} \Bigl ( \frac{q}{r_t} -q_0 \Bigr) \bigr]$. The derivation is given in \ref{The proof of the bound_second_chap6}.

%We apply the upper bound \eqref{bo1} for the proof.

By combining the inequalities~\eqref{Gamma_l_m_t_m_Delta2} and \eqref{G02}, we have
\begin{eqnarray}
\|\Gamma^{(q_n)}_{t_n} -\Gamma^{(q_0)}(t)  \| & \le  \sum_{m=1}^{n}\Delta (\Delta +1)^{m-1} \|\Gamma^{(q_0)}\|  \nonumber\\
& =\bigl[(\Delta +1)^{n}-1  \bigr] \cdot \|\Gamma^{(q_0)}\|  .\label{Delta_ineq_final1}
\end{eqnarray}
We can always find an operator $\Gamma^{(q_n)}_{t_n}$ such that $\|\Gamma^{(q_n)}_{t_n} -\Gamma^{(q_0)}(t)\|\le \|\Gamma^{(q_0)}\|$ (e.g. $\Gamma^{(q_n)}_{t_n}=0$), and hence we  only have to consider the range $(\Delta +1)^{n}-1   \le 1$ in the above inequality and obtain~\footnote{
Proof of \eqref{Delta_ineq_final2}: the inequality $(x+1)^{n}-1\le1$ is satisfied for $x\le 2^{1/n}-1$.
By the use of the fact that $(x+1)^{n}-1$ is the concave function for $x\ge0$, we have 
$
(x+1)^{n}-1 \le n x (x_0 +1)^{n-1}
$
for $0< x\le x_0$ with $x_0$ a positive constant. 
By choosing $x_0=2^{1/n}-1$, we obtain $(x+1)^{n}-1 \le n x 2^{(n-1)/n}$. $\square$ }
\begin{eqnarray}
(\Delta +1)^{n}-1 \le 2^{1-1/n} \cdot n \cdot \Delta < 2n \Delta.\label{Delta_ineq_final2}
\end{eqnarray}

The inequality~\eqref{Delta_ineq_final1} reduces to the inequality~\eqref{bound_chap6_second} due to the inequality~\eqref{Delta_ineq_final2} and the definitions of $\Delta$ and $n$.
This completes the proof of Theorem~\ref{Thm:exp_dynamics}. $\square$
%
%\footnote{In the case where we consider the imaginary time-evolution of $\Gamma^{(q_0)}(it)= e^{H t} \Gamma^{(q_0)} e^{-H t}$, the expansion diverges at a critical temperature $t_c$. The same thing is also observed when we consider the Lieb-Robinson bound by the imaginary-time evolution~\cite{}. }

%%%%%%%%%%%%%%%%%%%%%%%%%%%%%%%%%%%%%%%%%%%%%%%%%%%%%%%%%%%%%%%%%%%%%%%%%%%%%%%%%%%%%%%%%%%%%%%%%%%%%%%%%%%%%%%%%%%%%%%%%%%%%%%%%%%%%%%%%%%%%%%%%%%%%%%%%%%%%%%%%%%%%%%%%%%%%%%%%%

%{~}

\section{Several implications}
%%%%%%%%%%%%%%%%%%%%%%%%%%%%%%%%%%%%%%%%%%%%%%%%%%%%%%%%%%%%%%%%%%%%%%%%%%%%%%%%%%%%%%%%%%%%%%%%%%%%%%%%%%%%%%%%%%%%%%%%%%%%%%%%%%%%%%%%%%%%%%%%%%%%%%%%%%%%%%%%%%%%%%%%%%%%%%%%%%%%%%%%%%%%%%%%%%%%%%%%%%%%%%%%%%%%%%%%%%%%%%%%%%%%%%%%%%%%%%%%%%%%%%%%%%%%%%%%%%%%%%%%%%%%%%%%%%%%%%%%%%%%%%%%%%%%%%%%%%%%%%%%%%%%%%%%%%%%%%%%%%%%%%%%%%%%%%%%%%%%%%%%%%%%%%%%%%%%%%%%%%%%%%%%%%%%%%%%%%%%%%%%%%%%%%%%%%%%%%%%%%%%%%%%%%%%%%%%%%%%%%%%%%%%%%%%%%%%%%%%%%%%%

\subsection{Stability of the topological order}
We here prove that the topological order is stable after the time evolution over $t\lesssim  \log N$.
For the definition of the topological order, we follow the same discussion as in Ref~\cite{PhysRevLett.97.050401}.
The concept of the topological order is usually defined with respect to Hamiltonians rather than quantum states.
However, we have several common properties which the topological ordered phases always satisfy.

Slightly generalizing the definition in Ref.~\cite{PhysRevLett.97.050401}, 
we here define that a quantum state $\ket{\psi}$ exhibits the topological order if and only if there exists another quantum state $\ket{\tilde{\psi}}$ which satisfies
\begin{eqnarray}
\bra{\psi} \Gamma^{(q)}  \ket{\psi} = \bra{\tilde{\psi}} \Gamma^{(q)}  \ket{\tilde{\psi}} \quad {\rm and} \quad \bra{\psi} \Gamma^{(q)}  \ket{\tilde{\psi}}=0 \nonumber
\end{eqnarray}
for arbitrary $q$-local operators $\Gamma^{(q)}$ with $q=\orderof{N^{q}}$ and $q>0$. 
Here, the main difference from Ref.~\cite{PhysRevLett.97.050401} is 
that we apply generic $q$-local perturbation instead of the spatially local perturbation.
If the quantum state satisfies the property, the coherence between the two states $\ket{\psi}$ and $\ket{\tilde{\psi}}$ can never be broken by any kinds of local operators.
It is known that topologically ordered phases usually satisfy these conditions, for example the ground states of the Kitaev's toric code model~\cite{ref:Kitaev03-toric}.   
In Ref.~\cite{PhysRevLett.97.050401}, for short-range interacting systems, they proved by using the Lieb-Robinson bound that the topological order persists at least for $t\lesssim N^{1/D}$ ($D$: system dimension). This is contrast to the present case where we apply the new bound~\eqref{bound_chap6_second} and ensure
 the time of the stability for $t\lesssim  \log N$.

In considering the stability of the topological order, we define the topological order with error $(q,\epsilon_q)$ as 
\begin{eqnarray}
| \bra{\psi} \Gamma^{(q)}  \ket{\psi} - \bra{\tilde{\psi}} \Gamma^{(q)}  \ket{\tilde{\psi}} | \le \epsilon_q  \quad {\rm and} \quad |\bra{\psi} \Gamma^{(q)}  \ket{\tilde{\psi}}|\le \epsilon_q, \nonumber
\end{eqnarray} 
where $\|\Gamma^{(q)}\|=q$. 
For exactly topologically ordered states, we have $\epsilon_q=0$ for $q \le q_0 = \orderof{N^{\alpha}}$ with $\alpha >0$.
We will see that after small-time evolution the error $\epsilon_q$ can be small sub-exponentially with respect to the system size. 
In evaluating the error $(q,\epsilon_q)$ in the case of the $k$-local Hamiltonians, we apply the bound \eqref{bound_chap6_second} in Theorem~\ref{Thm:exp_dynamics} instead of the Lieb-Robinson bound.
From the similar discussions as in Ref.~\cite{PhysRevLett.97.050401}, we can obtain 
\begin{eqnarray}
\epsilon_{q}(t) \le 2 \lceil \kappa t \rceil  
\exp \biggl [ -\frac{1}{\xi} \Bigl ( \frac{q_0}{r_t} -q \Bigr) \biggr],  \nonumber
\end{eqnarray}
where we use Theorem~\ref{Thm:exp_dynamics} in evaluating $| \bra{\psi(t)} \Gamma^{(q)}  \ket{\psi(t)} - \bra{\tilde{\psi}(t)} \Gamma^{(q)}  \ket{\tilde{\psi}(t)} |=
| \bra{\psi} \Gamma^{(q)}(-t)  \ket{\psi} - \bra{\tilde{\psi}} \Gamma^{(q)}(-t)  \ket{\tilde{\psi}}|$.
If we consider $q \le q_0/[2r_t]$, the error $\epsilon_{q}$ is bounded from above by
$\orderof{t} \cdot  \exp \left[ -q_0 e^{-\orderof{t}}  \right]$.
This means that the error of the topological order is small exponentially with respect to $q_0$ as long as $t\lesssim \log q_0 = \orderof{\log N}$.

%{~}
%
\subsection{Dynamics of probability distribution of macroscopic observables}
We finally discuss how to connect our Theorem~\ref{Thm:exp_dynamics}  to observable quantities. 
As an example, we consider an upper bound for distribution function for extensive quantities $A$ such as
$A=\sum_{i=1} a_{i}$ with $\|a_i\|=1$ for $i\in \Lambda$.
Let us  consider a product state $\ket{\Prod}$ and its time evolution $e^{-iHt}\ket{\Prod}$. 
Initially, the spins are independent of each other and the distribution of $A$ for $\ket{\Prod}$ is concentrated with the standard deviation at most of $\orderof{\sqrt{N}}$ due to the Chernoff bound~\cite{chernoff1952measure}.
After the time evolution, the spins can couple with each other but still maintains the local independence approximately. 
Hence, we expect that the distribution of the operator $A$ should be still concentrated in $\ket{\Prod(t)}$. 
Indeed, we can prove the following theorem:

\begin{theorem}\label{Thm:exp_concent}
Let $\Pi^A_{\ge z}$ be a projection operator onto the subspace of the eigenvalues of $A$ which are in $[z,\infty)$.
Then, the spectrum of $A$ in $\ket{\Prod(t)}$ is exponentially concentrated as
\begin{eqnarray}
\|\Pi_{\ge \langle A\rangle + R}^A\ket{\Prod(t)}\| \le c_1 \cdot \exp\biggl(-\frac{R}{c_2r_t \sqrt{ t N}} \biggr), \label{concentration_spectrum_A_chap6}
\end{eqnarray}  
with $c_1$ and $c_2$ $\orderof{1}$ constants, where $\langle A\rangle$ is average value with respect to $\ket{\Prod(t)}$. 
\end{theorem}

After a short time $t=\orderof{1}$, the distribution of $A$ is still strongly concentrated with a standard deviation $e^{\orderof{t}} \sqrt{N}$.
Thus, in the state $\ket{\Prod(t)}$, spins are still locally independent of each other.
We note that this also implies no macroscopic entanglement in terms of the quantum Fisher information~\cite{shimizu2005detection,frowis2012}. 
We expect that the inequality~\eqref{concentration_spectrum_A_chap6} can be improved to the Gaussian form by using the recent technique~\cite{chernoff_Kuwahara}.

{~}\\
\textit{Proof of Theorem~\ref{Thm:exp_concent}}.
For the proof, we focus on the fact that the product state $\ket{\Prod}$ is given by ground state of a $1$-local Hamiltonian, say $H_p$: $H_p = \sum_{i=1}^N h_i$ with $\|h_i\|=1$ for $i\in \Lambda$ and $H_p\ket{\Prod}=-N \ket{\Prod}$.
Note that the spectral gap $\Delta E$ between the ground state and the first excited states for $H_p$ is $\orderof{1}$.
Then, the state $\ket{\Prod(t)}$ is also a gapped ground state of $H_p(t)=\sum_{i=1}^N h_i(t)$.
We now expand $H_p(t)$ by the use of $\Pi^A_{[x r_t,r_t + x r_t)}:=\Pi^A_x$:
\begin{eqnarray}
H_p (t) &= \sum_{x,x'}  \Pi^A_{x} H_p (t) \Pi^A_{x'}  \nonumber\\
&:=\sum_{x,x'}\sum_{s=1}^{d_x} \sum_{s'=1}^{d_{x'}}  \bra{x,s} H_p(t) \ket{x',s'}\ket{x,s}\bra{x',s'}, \label{Tight_binding}
\end{eqnarray}
where we denote $\Pi^A_{x} = \sum_{s=1}^{d_x} \ket{x,s}\bra{x,s}$ with $d_x = \rank (\Pi^A_{x})$.

By applying Theorem~\ref{Thm:exp_dynamics},  we obtain 
\begin{eqnarray}
\left\|\Pi^A_{x} H_p (t) \Pi^A_{x'}\right\|\le C_ve^{-\mu(t)|x-x'|}
\label{t:eff}
\end{eqnarray}
with 
\begin{eqnarray}
C_v =  8N e^{5/(2\xi )}\lceil \kappa t \rceil   , \quad \mu = 1/ (2\xi),  
\end{eqnarray}
where the derivation is given in \ref{Chap6_calculation_of_Delta}.
This way, we can formally regard the Hamiltonian~\eqref{Tight_binding} as a tight-binding Hamiltonian; the position $x$ corresponds to the eigenvalue of $A$.
Remembering that the state $\ket{\Prod(t)}$ is the gapped ground state of $H_p(t)$, 
the distribution of the position $x$ should be localized due to the spectral gap $\Delta E$~\cite{o1973exponential,kuwahara2013upper,1742-5468-2016-5-053103}; from Ref~\cite{1742-5468-2016-5-053103}, the localization length is proportional to $\sqrt{C_v \mu^3/\Delta E} \propto \sqrt{tN}$ with $\Delta E=\orderof{1}$.
%\begin{eqnarray}
%\sqrt{4 C_v\sum_{x=0}^\infty(x+1)^2 e^{-\mu x}}  \propto \sqrt{tN}.
%\end{eqnarray}  
Now, $1/r_t$ times of the eigenvalue of $A$ corresponds to the position $x$, 
which yields the localization length of $A$ which is smaller than $\orderof{r_t \sqrt{tN}}$.
This complete the proof.
$\square$

Before closing this subsection, we discuss the relationship to the spin squeezing~\cite{Ma201189}, where the magnetization along a certain axis can be squeezed 
by broadening the variance along another axis. The squeezing protocols contains the process to create large fluctuation and hence Theorem~\ref{Thm:exp_concent} is applicable to the estimation of necessary time for the squeezing creation; for example, let $M_{x}$, $M_{y}$, and $M_{z}$ be
the usual collective spin operators and assume that the total spin amplitude is $\orderof{N}$. Then, the uncertainty principle 
ensures $\Delta M_{x}\Delta M_{y} \ge C N$ with $C$ a constant of $\orderof{1}$ and $\Delta(\cdot)$ denoting the fluctuation. Theorem~\ref{Thm:exp_concent} implies that the necessary time for the squeezing of 
$\Delta M_{x}/\Delta M_{y}=\xi_{\rm sq}<1$ (i.e., $\Delta M_{y} \ge\sqrt{CN/\xi_{\rm sq}}$) is at least proportional to 
$\log (\xi_{\rm sq}^{-1})$. This time scale is apparently not consistent to the many previous works~\cite{Ma201189}, 
in which the necessary time is $\orderof{1}$ even if $\xi_{\rm sq}=N^{-\alpha}$ with $\alpha >0$ for protocols using two-body all-to-all interactions like Lipkin-Meshkov-Glick model~\cite{PhysRevA.91.053612}.  
The key point is that we now assume \textit{the extensiveness of the Hamiltonian}~\eqref{g-extensive}, whereas in spin squeezing literatures, the Hamiltonians are super extensive. By taking this point into account, we can resolve the inconsistency.

%%%%%%%%%%%%%%%%%%%%%%%%%%
%%%%%%%%%%%%%%%%%%%%%%%%%%%%%%%%%%%%%%%%%%%%%%%%%%%%%%%%%%%%%%%%%%%%%%%%%%%%%%%%%%%%%%%%%%%%%%%%%%%%%%%%%%%%%%%%%%%%%%%%%%%%%%%%%%%%%%%%%%%%%%%%%%%%%%%%%%%%%%%%%%%%%%%%%%%%%%%%%%

\section{Outlook}
We have given a new bound on the quantum dynamics which are governed by the $k$-local Hamiltonian and shown some applications which have not be derived from 
the Lieb-Robinson bound for long-range interacting systems.
Our main theorem characterizes the number of spins which cause many-body quantum effect due to time evolution.

As further applications, it is one of the most important problems whether we can apply the present results to the quasi-adiabatic continuation~\cite{ref:Hastings-locality,PhysRevB.72.045141} for the $k$-local Hamiltonians. 
%In more detail, let us consider an evolution due to the change of an internal parameter of the Hamiltonian instead of the time evolution. 
In more detail, let the Hamiltonian be $H(s) = H +sV$ with $0\le s \le 1$, where $H$ and $V$ are $k$-local Hamiltonians, respectively.
Here we assume that the ground state of $H(s)$, say $\ket{E_0(s)}$, is non-degenerate and is separated from the excited states by a non-vanishing gap for $0\le s \le 1$.
Then, the evolution of the state can be described similarly to the time evolution:
$
\frac{d}{ds} \ket{E_0(s)} = i D(s)  \ket{E_0(s)}, \label{Adi_cont_chap6}
$
where $D(s)$ is defined as the adiabatic continuation operator. 
We therefore can conclude that the parameter evolution of $\ket{E_0(s)}$ is formally equivalent to the time evolution by $D(s)$. 
Our interest is whether the operator $D(s)$ can be approximated by $k$-local operators. 
In that case, we can analyze the perturbative effect to the ground state by the use of Theorem~\ref{Thm:exp_dynamics}, 
from which we can generalize the stability analysis for the topological order in the short-range interacting systems~\cite{bravyi2010topological,
bravyi2011short,michalakis2013stability}.

Another interesting direction is to identify an efficient description of the state $\ket{\Prod(t)}$,
for example, by the use of the tensor network state~\cite{perez2006matrix}.
Because the multipartite effect is highly suppressed as long as $t=\orderof{1}$, 
we expect that the small time-evolution can be efficiently simulated. 
Moreover, if the Hamiltonian contains randomness in its interaction, 
the locality of the multipartite coupling might persistently maintain beyond the time scale $t\gtrsim \log N$, in the similar manner as the short-range interacting systems~\cite{PhysRevLett.99.167201}, where the entanglement growth is logarithmically slow.

Finally, can we observe our bound experimentally?
In particular, Ref.~\cite{richerme2014non} has demonstrated the Lieb-Robinson bound in long-range interacting systems.
Our new bound in Theorem~\ref{Thm:exp_dynamics} can be in principle observed in the same experimental setup, for example, 
by looking at the probability distribution which follows Theorem~\ref{Thm:exp_concent}.

\section*{ACKNOWLEDGMENT}

We are grateful to Naomichi Hatano, Tatsuhiko Shirai, Takashi Mori and Kaoru Yamamoto for helpful discussions and comments on related topics.  
This work was partially supported by the Program for Leading Graduate Schools (Frontiers of Mathematical Sciences and Physics, or FMSP), MEXT, Japan. 
The author was also supported by World Premier International Research Center Initiative (WPI), Mext, Japan. 
Finally, TK acknowledges the support from JSPS grant no. 2611111.

\bibliographystyle{iopart-num}

{~}

\bibliography{Exp_dynamics}

\appendix

\section{Proof of Theorem~\ref{k_local_fund}}\label{Appendix_A}
We show the proof of Theorem~\ref{k_local_fund}. 
For the proof, we take the following two steps.

{\bf (Step 1)}  
We first consider a class of the commuting Hamiltonians $H^\co$ as 
\begin{eqnarray}
H^{\co}=\sum_{|X|\le k} h_{X}\quad {\rm with} \quad [h_X,h_{X'}]=0  ,  \quad \forall X,X',
\label{Extensive_Hamiltonian_chap_co6}
\end{eqnarray}  
and prove the inequality 
  \begin{eqnarray}
\| [H^\co ,\Gamma^{(q)}]\| \le 6gq  \|\Gamma^{(q)}\| . 
\label{Basic_inequality_for_Hco_local}
\end{eqnarray}

{\bf (Step 2)} 
Secondly, we prove that any $k$-local $g$-extensive Hamiltonian can be decomposed into the sum of \textit{commuting} Hamiltonians:
\begin{eqnarray}
H=\frac{1}{\bar{n}} \sum_{m=1}^{\bar{n}} H_m^\co, 
\label{SC_decomp}
\end{eqnarray}  
where each of $\{H_m^\co\}$ is $k$-local and $(gk)$-extensive.

After these two steps, we can prove Theorem~\ref{k_local_fund} as  
  \begin{eqnarray}
\| [H ,\Gamma^{(q_0)}]\| &\le \frac{1}{\bar{n}} \sum_{m=1}^{\bar{n}} \| [H_m^\co ,\Gamma^{(q)}]\| \nonumber\\
&\le \frac{1}{\bar{n}} \sum_{m=1}^{\bar{n}} 6 gk q \|\Gamma^{(q)} \| = 6 gk q \|\Gamma^{(q)} \|.
\end{eqnarray}
This proves Theorem~\ref{k_local_fund}.
In the following subsections, we will prove the statements in the Steps 1 and 2. 

\subsection{Step 1} 
We here prove the upper bound \eqref{Basic_inequality_for_Hco_local} for $\|[H^{\co}, \Gamma^{(q_0)}]\|$.
We first decompose $H^\co$ as follows:
 \begin{eqnarray}
H^\co= H^{'\co}+  \delta H^\co , \nonumber
\end{eqnarray}
where 
 \begin{eqnarray}
H^{'\co} := \sum_{j=-\infty}^{\infty}\epsilon ( j+1/2)  \Pi_{[\epsilon j,\epsilon  j+\epsilon )}, \quad \delta H^\co := H^\co- H^{'\co} \nonumber
\end{eqnarray}
and $\Pi_{[\epsilon j,\epsilon  j+\epsilon )}$ is a projection operator onto the eigenspace of $H^{\co}$ with the eigenvalues $[\epsilon j,\epsilon  j+\epsilon )$.
We set the value of $\epsilon$ afterward.
Note that the operator $\Pi_{[\epsilon j,\epsilon  j+\epsilon )}$ may be the null operator.
From the definition, we have 
 \begin{eqnarray}
\| \delta H^\co \| \le \frac{\epsilon}{2}. \nonumber
\end{eqnarray}
We then obtain
 \begin{eqnarray}
\|[\Gamma^{(q)}, H^\co]\| &= \|[\Gamma^{(q)},  H^{'\co}+\delta H^\co]\|  \le \|[\Gamma^{(q)},  H^{'\co}]\| + \|[\Gamma^{(q)},  \delta H^\co]\|  ,\label{Theo_13_total_H_co} 
\end{eqnarray}
which necessitates that we calculate $\|[\Gamma^{(q)},  H^{'\co}]\|$ and $\|[\Gamma^{(q)},  \delta H^\co]\|$ separately.

We first obtain the norm of $[\Gamma^{(q)},  \delta H^{\co}]$ as follows:
 \begin{eqnarray}
\|[\Gamma^{(q)},  \delta H^\co]\| \le 2 \|\Gamma^{(q)}\| \cdot \| \delta H^\co\| \le  \epsilon  \|\Gamma^{(q)}\|. \label{Theo_13_delta_H_co}
\end{eqnarray}
We second obtain
 \begin{eqnarray}
[\Gamma^{(q)}, H^{'\co}]&=\sum_{ j, j'} \Pi_{[\epsilon j,\epsilon j+\epsilon )} ( \Gamma^{(q)} H^{'\co}   -H^{'\co} \Gamma^{(q)}) \Pi_{[\epsilon j',\epsilon j'+\epsilon )}  \nonumber\\
&= \sum_{j,j'} \epsilon (j'-j)\Pi_{[\epsilon j,\epsilon  j+\epsilon )} \Gamma^{(q)} \Pi_{[\epsilon j',\epsilon j'+\epsilon )} .
\label{Equality_sc_Hamiltonian_commu1}
\end{eqnarray}
Because we can obtain the norm of $[\Gamma^{(q)}, H^{'\co}]$ from the equality  
\begin{eqnarray}
\|[\Gamma^{(q)}, H^{'\co} ]\| = \max_{\ket{\psi}} |\bra{\psi}[\Gamma^{(q)}, H^{'\co}] \ket{\psi}|,\nonumber
\end{eqnarray} 
we, in the following, calculate the upper bound of $|\bra{\psi}[\Gamma^{(q)}, H^{'\co}] \ket{\psi}|$ for arbitrary quantum states $\ket{\psi}$.
From Eq.~\eqref{Equality_sc_Hamiltonian_commu1}, we have
 \begin{eqnarray}
\fl |\bra{\psi}[\Gamma^{(q)}, H^{'\co}] \ket{\psi}| 
&=\biggl| \sum_{j,j'}\epsilon (j'-j)  \bra{\psi} \Pi_{[\epsilon j,\epsilon  j+\epsilon )}\Gamma^{(q)} \Pi_{[\epsilon j',\epsilon j'+\epsilon )} \ket{\psi} \biggr| \nonumber\\
\fl &\le \sum_{j,j'}\epsilon  |j'-j|\cdot  \| \bra{\psi}\Pi_{[\epsilon j,\epsilon  j+\epsilon )}\| \cdot \| \Pi_{[\epsilon j,\epsilon  j+\epsilon )}  \Gamma^{(q)} \Pi_{[\epsilon j',\epsilon j'+\epsilon )}\|\cdot  \| \Pi_{[\epsilon j',\epsilon j'+\epsilon )} \ket{\psi} \| \nonumber\\
\fl &=: \epsilon \sum_{j,j'} |j'-j|  \alpha_{j'}\alpha_{j}   \Gamma_{j,j'}^{(q)} , \nonumber
\end{eqnarray}
where $\alpha_{j} := \|\Pi_{[\epsilon j,\epsilon  j+\epsilon )} \ket{\psi}\|$ and $\Gamma_{j,j'}^{(q)}:=  \| \Pi_{[\epsilon j,\epsilon  j+\epsilon )}  \Gamma^{(q)} \Pi_{[\epsilon j',\epsilon j'+\epsilon )}\| $; note that $\sum_{j} \alpha_j^2=1$.

Because we are now considering the commuting Hamiltonian, due to $g$-extensiveness~\eqref{g-extensive}, we have~\cite{ref:Arad-kuwahara,kuwahara2015local}
 \begin{eqnarray}
 &\| \Pi_{[E',\infty)}\Gamma^{(q)} \Pi_{(-\infty,E]} \| \le \|\Gamma^{(q)}\| \for |E'-E| \le 2 g q, \nonumber\\
& \| \Pi_{[E',\infty)} \Gamma^{(q)} \Pi_{(-\infty,E]} \| =0 \for |E'-E| > 2 g q,
\end{eqnarray}
where the first inequality comes from the trivial bound of $\| \Pi_{[E',\infty)}\Gamma^{(q)} \Pi_{(-\infty,E]} \| \le \| \Pi_{[E',\infty)}\|\cdot \| \Gamma^{(q)}\| \cdot \| \Pi_{(-\infty,E]} \|=  \| \Gamma^{(q)}\|$.
This gives the following inequality:
 \begin{eqnarray}
 &\Gamma_{j,j'}^{(q)}  \le \|\Gamma^{(q)}\| \for |j'-j| \le 1 + \frac{2 g q}{\epsilon}, \nonumber\\
&\Gamma_{j,j'}^{(q)}  =0 \for  |j'-j| > 1 + \frac{2 g q}{\epsilon}  .\label{com_lemma_o_i_j}
\end{eqnarray}
Because of the inequality~\eqref{com_lemma_o_i_j}, we have
 \begin{eqnarray} 
|\bra{\psi}[\Gamma^{(q)}, H^\co] \ket{\psi}| &=\epsilon \sum_{|j'-j| \le 1 + 2 g q/\epsilon} |j'-j|  \alpha_{j'}\alpha_{j} \Gamma_{j,j'}^{(q)}\nonumber\\
&\le \epsilon\|\Gamma^{(q)}\|\sum_{|j'-j| \le 1 + 2 g q/\epsilon} |j'-j|  \frac{\alpha_{j'}^2+\alpha_{j}^2}{2}   \nonumber\\
&\le \epsilon\|\Gamma^{(q)}\|\sum_{|j'-j| \le 1 + 2 g q/\epsilon} |j'-j|  \alpha_{j}^2 \nonumber\\
&\le \epsilon \|\Gamma^{(q)}\|  \sum_{j'=-\lfloor 1 + \frac{2 g q}{\epsilon}  \rfloor}^{ \lfloor 1 + \frac{2 g q}{\epsilon}  \rfloor }|j'|  \sum_j \alpha_{j}^2 
\nonumber\\
 &\le\epsilon\|\Gamma^{(q)}\| \Bigl \lfloor 1 + \frac{2 g q}{\epsilon} \Bigr \rfloor  \biggl (\Bigl \lfloor 1 + \frac{2 g q}{\epsilon} \Bigr \rfloor +1 \biggr), \label{Theom_13_tilde_H_co}
\end{eqnarray}
where $\lfloor\cdots \rfloor$ denotes the floor function.

The inequalities~\eqref{Theo_13_delta_H_co} and \eqref{Theom_13_tilde_H_co} reduce the inequality~\eqref{Theo_13_total_H_co} to
 \begin{eqnarray}
\|[\Gamma^{(q)}, H^\co]\|  \le\epsilon\|\Gamma^{(q)}\| + \epsilon\|\Gamma^{(q)}\| \Bigl \lfloor 1 + \frac{2 g q}{\epsilon} \Bigr \rfloor  \biggl (\Bigl \lfloor 1 + \frac{2 g q}{\epsilon} \Bigr \rfloor +1 \biggr). \nonumber
\end{eqnarray}
We here choose $\epsilon =2 g q + \delta\epsilon$ ($\delta \epsilon >0$) and obtain
 \begin{eqnarray}
\|[\Gamma^{(q_0)}, H^\co]\|\le (6 g q+ 3 \delta\epsilon) \|\Gamma^{(q_0)}\|. \nonumber
\end{eqnarray}
By taking the limit of $\delta \epsilon \to +0$, we finally obtain $\|[\Gamma^{(q)}, H^\co]\| \le  6g q \|\Gamma^{(q_0)}\|$.

\subsection{Step 2} 
We now prove the existence of the decomposition~\eqref{SC_decomp}.
For the proof, we first introduce a parameter $\epsilon$ and define unit operators $\{\tilde{h}_X\}$ as follows:
\begin{eqnarray}
\tilde{h}_X := \epsilon \frac{h_X}{\|h_X\|} ,
\end{eqnarray}
where $\{h_X\}$ are components of the Hamiltonian.
By the use of  $\{\tilde{h}_X\}$, we can define the `discretized' Hamiltonian $H'$ as
\begin{eqnarray}
H' = \sum_X  N_{h_X} \tilde{h}_X \quad {\rm with } \quad N_{h_X} := \biggl\lfloor \frac{\|h_X\|}{\epsilon} \biggr\rfloor   .
\end{eqnarray}
Note that we have $\|H'-H\| =\orderof{\epsilon N}$, which vanishes in the limit of $\epsilon \to 0$.
Because of the extensiveness of the Hamiltonian, the number of unit operators $\{\tilde{h}_X\}$ which one spin can contain should be bounded from above by
\begin{eqnarray}
\sum_{X\ni i}  N_{h_X} \le \frac{g}{\epsilon}  \label{number_unit_ss}
\end{eqnarray}

We, in the following, decompose the Hamiltonian $H'$ into $\bar{n}$ commuting Hamiltonians $\{H^\co_m\}_{m=1}^{\bar{n}}$ by the use of $\{\tilde{h}_X\}$, where $\bar{n}$ is an integer.
We here correspond a set of $\{\tilde{h}_{X_m^{(i)}}\}_{i=1}^{N_m}$ to one commuting Hamiltonians $H_m^\co$, namely
\begin{eqnarray}
&H_m^{\co} = \tilde{h}_{X_m^{(1)}} + \tilde{h}_{X_m^{(2)}} + \cdots + \tilde{h}_{X_m^{(N_m)}} ,\nonumber\\
& {\rm s.t.} \quad X_m^{(j)} \cap X_m^{(k)} =0 \for \forall j,k  \label{construction_Commuting_Hm}
\end{eqnarray}
Note that the Hamiltonian $H^{\co}$ is commuting, $k$-local and $\epsilon$-extensive because of $\|\tilde{h}_{X_m^{(i)}}\|=\epsilon$.
Hence, if we can decompose the total Hamiltonian $H'$ with $\bar{n}=k\lfloor g/\epsilon \rfloor$, we obtain
 \begin{eqnarray}
H'=\frac{1}{\bar{n}}\sum_{m=1}^{\bar{n}} \bar{n} H_m^{\co}.
\end{eqnarray}
The commuting Hamiltonians $\{\bar{n} H_m^{\co}\}_{m=1}^{\bar{n}}$ are $k$-local and $(gk)$-extensive.
Thus, by taking $\epsilon\to 0$, we have $\|H'- H\|\to 0$ and this completes the proof of the decomposition~\eqref{SC_decomp}.

In the following, we will prove that $\bar{n}=k\lfloor g/\epsilon \rfloor$ is a sufficient number of commuting Hamiltonians $\{H^\co_m\}$ to construct $H'$.
According to the definition of $H_m^\co$ in Eq.~\eqref{construction_Commuting_Hm}, we denote the support of $H_m^\co$ by $L_m$, namely 
\begin{eqnarray}
L_m =X_m^{(1)} \cup X_m^{(2)} \cup \cdots  \cup X_m^{(N_m)}.
\end{eqnarray}
In the construction of $\{H^\co_m\}$, we decompose the Hamiltonian $H'$ such that $|L_1|\le |L_2|\le \cdots \le |L_{\bar{n}}|$.

We first collect the units $\{\tilde{h}_X\}$ for $H^\co_1$ so that $|L_1|$ may be as large as possible, mathematically,
\begin{eqnarray}
H'_{L_1^\co} := \sum_{X:X\in L_1^\co }   N_{h_X} \tilde{h}_X = 0 .
\end{eqnarray}
This means that there are no unit Hamiltonians $\{\tilde{h}_X\}$ outside of $L_1$.
Second, we also collect $\{\tilde{h}_X\}$ into $H^\co_2$ so that $|L_2|$ may be as large as possible:
\begin{eqnarray}
H^{(2)}_{L_2^\co} =0  \quad {\rm with} \quad H^{(2)}:= H'- H^\co_{1}.
\end{eqnarray}
By repeating this process, we construct the commuting Hamiltonians $\{H^\co_m\}$ so that they may satisfy
\begin{eqnarray}
H^{(m)}_{L_m^\co} =0  \quad {\rm with} \quad H^{(m)}:= H'- \sum_{j=1}^{m-1} H^\co_{j}. \label{Cond_for_H_m^c}
\end{eqnarray}
It means that  there are no unit Hamiltonians $\{\tilde{h}_X\}$ outside of $L_m$ in considering $H^{(m)}$.
Note that if $L_{\bar{n}+1}=0$ we also have $H^{(\bar{n}+1)}= H'- \sum_{j=1}^{\bar{n}} H^\co_{j}=0$ and the complete decomposition has been achieved.
In the following, we have to prove $L_{\bar{n}+1}=0$ with $\bar{n}=k \lfloor g /\epsilon \rfloor$.

For the proof, we assume $L_{\bar{n}+1}\neq \emptyset$, or equivalently $H^{\co}_{\bar{n}+1}\neq \hat{0}$, and prove the contradiction. 
We use the fact that any terms $\{ \tilde{h}_{X_{\bar{n}+1}^{(j)}} \}_{j=1}^{N_{\bar{n}+1}}$ supported in $L_{\bar{n}+1}$ should satisfy
\begin{eqnarray}
X_{\bar{n}+1}^{(j)}  \cap L_{m}\neq \emptyset \label{Common_subspace}
\end{eqnarray}
for $m=1,2,3,\ldots, \bar{n}$, which comes from the condition \eqref{Cond_for_H_m^c}.
Because of \eqref{Common_subspace} and $|X_{\bar{n}+1}^{(j)}|\le k$, 
at least $(\bar{n}/k)$ subspaces in $\{L_m\}_{m=1}^{\bar{n}}$ have a common support; 
for example, if $|X_{\bar{n}+1}^{(j)} |=1$ or $X_{\bar{n}+1}^{(j)}$ contains only one spin (e.g. spin $i$), the relation \eqref{Common_subspace} means that all of $\{L_m\}_{m=1}^{\bar{n}}$ contain the spin $i$.

Therefore, there exists a set $\{L_{m_i}\}_{i=1}^{\bar{n}/k}$ such that 
\begin{eqnarray}
X_{\bar{n}+1}^{(j)}  \cap L_{m_1} \cap L_{m_2}  \cap L_{m_3}  \cap \cdots \cap L_{m_{\bar{n}/k}}  \neq \emptyset.
\end{eqnarray}
We denote this common support by $\tilde{L}$.
Then, the spins in $\tilde{L}$ should be contained in all the Hamiltonians $\{H_{m_i}^\co\}_{m=1}^{\bar{n}/k}$ and $H_{\bar{n}+1}^\co$, whereas,  due to the inequality \eqref{number_unit_ss},  one spin can contains up to $(\bar{n}/k)$ unit operators $\{\tilde{h}_X\}$; note that $\lfloor g/\epsilon \rfloor=\bar{n}/k$.
Thus, we prove the contradiction.  $\square$

\section{Bound~(\ref{bo1}) for a small-time evolution} \label{Bound for a small-time evolution}

We here prove the inequality~\eqref{bo1} for a small-time evolution:
in order to obtain the bound, we expand $e^{-i H t} \Gamma^{(q_0)} e^{i H t}$ by the Hadamard lemma as in Eq.~\eqref{Hadmal_expansion_expd2},
which we reproduce here:
 \begin{eqnarray}
e^{-iH t} \Gamma^{(q_0)} e^{iH t}  = \sum_{m=0}^\infty \frac{(-i t)^m}{m!} L_m,  \label{Hadmal_expansion_expd2_again}
\end{eqnarray}
where $q_0=\Gamma^{(q_0)}$ and $L_m =\overbrace{ [H,[H, \cdots [H}^m ,\Gamma^{(q_0)}]]\cdots ]$. 
This expansion can be terminated if the expansion converges rapidly as $m$ increases. 
The termination at $m=m_0$ gives the local approximation  of the operator $\Gamma^{(q_0)}(t)$, namely
 \begin{eqnarray}
\Gamma^{(q)}= \sum_{m=0}^{m_0} \frac{(i t)^m}{m!} L_m,  \label{appro_m_0_expansion}
\end{eqnarray}
Because $L_m$ is at most $(q_0+k m_0)$-local,  in order to make the operator~\eqref{appro_m_0_expansion} less than or equal to $q$-local, we take
 \begin{eqnarray}
m_0 = \biggl\lfloor\frac{q-q_0}{k} \biggr\rfloor. \nonumber
\end{eqnarray}

Our purpose now  is  to calculate the error due to  the cutoff of the above expansion, namely
 \begin{eqnarray}
\biggl\| \sum_{m=m_0+1}^\infty \frac{(- i t)^m}{m!} L_m \biggl\| \le \sum_{m=m_0+1}^\infty \frac{t^m}{m!} \|L_m\|  ,  \label{Hadmal_expansion2_expd2}
\end{eqnarray}
Now, we can apply Theorem~\ref{k_local_fund} to evaluate $\|L_m\|$:
 \begin{eqnarray}
\|L_m\| &=\| \overbrace{ [H,[H, \cdots [H}^m ,\Gamma^{(q_0)}]]\cdots ]  \| \nonumber\\
&\le \lambda^m \frac{q_0}{k}\frac{q_0+k}{k} \frac{q_0+2k}{k} \cdots \frac{q_0+(m-1)k}{k}\|\Gamma^{(q_0)}\|   \nonumber\\
&= \lambda^m \frac{(\lceil r \rceil + m-1)!}{(\lceil r \rceil-1)!} \|\Gamma^{(q_0)}\| \le  2^{r+m} m!\lambda^m \|\Gamma^{(q_0)}\|, \nonumber
\end{eqnarray}
where we define $r:= q_0/k$ and ${}_{\lceil r \rceil + m-1} C_m \le 2^{\lceil r \rceil + m-1}\le 2^{r+ m}$.
We thus obtain
 \begin{eqnarray}
\fl \sum_{m=m_0+1}^\infty \frac{t^m}{m!} \|L_m\|& \le 2^{r}\sum_{m=m_0+1}^\infty (2\lambda t)^m\|\Gamma^{(q_0)}\| \le 2^{q/k }\frac{(2\lambda t)^{(q-q_0)/k}}{1-2\lambda t}\|\Gamma^{(q_0)}\| ,
\end{eqnarray}
for $t\le (2\lambda)^{-1}$, where we used the inequality $m_0+1 \ge (q-q_0)/k$ in the last inequality.
By replacing $2\lambda=\kappa/2$, we can obtain the inequality~\eqref{bo1}.
%

%%%%%%%%%%%%%%%%%%%%%%%%%%%%%%%%%%%%%%%%%%%%%%%%%%%%%%%%%%%%%%%%%%%%%%%%%%%%%%%%%%%%%%%%%%%%%%%%%%%%%%%%%%%%%%%%%%%%%%%%%%%%%%%%%%%%%%%%%%%%%%%%%%%%%%%%%%%%%%%%%%%%%%%%%%%%%%%%%%

\section{The proof of the inequality~(\ref{G02})} \label{The proof of the bound_second_chap6}

%We start from the inequality~\eqref{Gamma_l_m_t_m_Delta2}.
We here prove the existence of the set $\{q_m, \Gamma^{(q_m)}_{t_m}\}_{m=1}^n$ which satisfies the inequality~\eqref{G02}.
Because of $\delta t \le 1/\kappa $, we can apply the bound~\eqref{bo1} to estimate the norm $\bigl \| \Gamma^{(q_m)}_{t_m} -  \Gamma^{(q_{m-1})}_{t_{m-1}} (\delta t)   \bigr\|$. 
As shown in the next subsection, we can find a set  $\{q_m, \Gamma^{(q_m)}_{t_m}\}_{m=1}^n$ such that
\begin{eqnarray}
\bigl \| \Gamma^{(q_m)}_{t_m} -  \Gamma^{(q_{m-1})}_{t_{m-1}} (\delta t)   \bigr\|  \le \|\Gamma^{(q_{m-1})}_{t_{m-1}}  \| \Delta \label{Gamma0_l_m_t_m_Delta0}
\end{eqnarray}
for $m=1,2,\ldots,n$, respectively, where 
\begin{eqnarray}
\Delta := 4 \exp \left [ -\frac{1}{\xi} \left ( \frac{q}{r_t} -q_0 \right) \right] . \label{Def_Delta_Sup}
\end{eqnarray}

By the use of the inequality~\eqref{Gamma0_l_m_t_m_Delta0}, we can obtain the following inequality:
\begin{eqnarray}
 \|\Gamma^{(q_m)}_{t_m}\| \le (\Delta +1)^m\|\Gamma^{(q_0)}\|. \label{Gamma_l_m_t_m_Delta}
\end{eqnarray}
We prove this inequality by the induction method. 
For $m=1$, we have
\begin{eqnarray}
\fl \|\Gamma^{(q_1)}_{t_1}\| &= \|\Gamma^{(q_1)}_{t_1} - \Gamma^{(q_0)}_{t_0} + \Gamma^{(q_0)}_{t_0} \| 
\le  \|\Gamma^{(q_1)}_{t_1} - \Gamma^{(q_0)}_{t_0} \| +\|\Gamma^{(q_0)}_{t_0} \|  \le (\Delta +1)\|\Gamma^{(q_0)}\|,\nonumber
\end{eqnarray}
where the last inequality came from \eqref{Gamma0_l_m_t_m_Delta0} and we used the definition of  $ \Gamma^{(q_{0})}_{t_{0}} = \Gamma^{(q_0)}$.
We then assume the inequality~\eqref{Gamma_l_m_t_m_Delta} for $m\le m_0$ and prove it for $m=m_0+1$ as follows:
\begin{eqnarray}
\|\Gamma^{(q_{m_0+1})}_{t_{m_0+1}}\| &= \|\Gamma^{(q_{m_0+1})}_{t_{m_0+1}}- \Gamma^{(q_{m_0})}_{t_{m_0}} +\Gamma^{(q_{m_0})}_{t_{m_0}}\|  \nonumber\\
&\le( \Delta+1) \|\Gamma^{(q_{m_0})}_{t_{m_0}} \| \le (\Delta +1)^{m_0+1}\|\Gamma^{(q_0)}\|.\nonumber
\end{eqnarray}
This completes the proof of the inequality~\eqref{Gamma_l_m_t_m_Delta}.

By combining the inequalities~\eqref{Gamma0_l_m_t_m_Delta0} and \eqref{Gamma_l_m_t_m_Delta}, we prove the inequality~\eqref{G02}.

\subsection{Proof of the inequality~(\ref{Gamma0_l_m_t_m_Delta0})} \label{Chap6_calculation_of_Delta}

We first calculate $\bigl \| \Gamma^{(q_m)}_{t_m} -  \Gamma^{(q_{m-1})}_{t_{m-1}} (\delta t) \bigr\| $ for some $q_m$ and $q_{m-1}$.
Because of the upper bound~\eqref{bo1} for a small-time evolution, there exists an operator $\Gamma^{(q_m)}_{t_m}$ for any $\Gamma^{(q_{m-1})}_{t_{m-1}}$ such that
\begin{eqnarray}
\bigl \|\Gamma^{(q_m)}_{t_m} -  \Gamma^{(q_{m-1})}_{t_{m-1}} (\delta t)  \bigr\|  &\le 
2^{q_{m-1}/k}\frac{(\kappa\delta t/2)^{(q_m-q_{m-1})/k} }{1-\kappa \delta t/2}  \bigl\| \Gamma^{(q_{m-1})}_{t_{m-1}} \bigr\| \nonumber\\
&\le 2^{\frac{2q_{m-1}-q_m}{k}+1}  \bigl\| \Gamma^{(q_{m-1})}_{t_{m-1}} \bigr\| , \label{Calculation_Delta_1_sub}
\end{eqnarray}
where the definition~\eqref{Definition_delta_t_chap6} give $\kappa \delta t/2 \le 1/2$ and in the last inequality, we use the fact  that the function $x^{(q_m-q_{m-1})/k}/(1-x)$  monotonically increases for $0\le x<1$.

We now define a positive integer $\delta_q$ such that
\begin{eqnarray}
q_m = 2q_{m-1} + \delta_q \nonumber
\end{eqnarray}
for $m=1,2,\ldots,n$.
We then obtain 
 \begin{eqnarray}
q_n = 2^n (q_{0} + \delta_q ) -\delta_q. \nonumber
\end{eqnarray}
Because of the condition $q_n \le q$, we have to take $\delta_q$ so that it may satisfy the inequality 
\begin{eqnarray}
2^n (q_{0} + \delta_q ) -\delta_q \le q , \quad {\rm or} \quad 
\delta_q \le \frac{q-2^n q_0}{2^n-1}. \nonumber
\end{eqnarray}
Based on this inequality, we choose $\delta_q$ as $\lfloor  (q-2^n q_0)/(2^n-1) \rfloor $.
By combining the inequality~\eqref{Calculation_Delta_1_sub} with the definition $\delta_q:= q_m - 2q_{m-1}$,  we finally obtain
\begin{eqnarray}
\bigl \|\Gamma^{(q_m)}_{t_m} -  \Gamma^{(q_{m-1})}_{t_{m-1}} (\delta t)  \bigr\| &\le 2^{-\frac{\delta_q}{k}+1}  \bigl\| \Gamma^{(q_{m-1})}_{t_{m-1}} \bigr\|  \nonumber\\
&\le  4\| \Gamma^{(x_{m-1})}_{(m-1)t}\| \exp \biggl ( - \log 2 \frac{q/(2^n-1) - q_0}{k} \biggr). \label{Gamma0_l_m_t_m_Delta0_for1}
\end{eqnarray}
We here notice the equality 
\begin{eqnarray}
\Delta&= 4 \exp \biggl [ -\frac{1}{\xi} \Bigl ( \frac{q}{r_t} -q_0 \Bigr) \biggr]=4 \exp \left ( - \log 2 \frac{q/(2^n-1) - q_0}{k} \right),\label{Gamma0_l_m_t_m_Delta0_for2}
\end{eqnarray}
because of the definitions of $\xi$, $r_t$ and $n$ as in Theorem~\ref{Thm:exp_dynamics} and Eq.~\eqref{Definition_delta_t_chap6}.
We thus prove the inequality~\eqref{Gamma0_l_m_t_m_Delta0} from \eqref{Gamma0_l_m_t_m_Delta0_for1} and \eqref{Gamma0_l_m_t_m_Delta0_for2}. 

\section{Derivation of the inequality~(\ref{t:eff})}
In this section, we consider the norm of
\begin{eqnarray}
\left\| \Pi^A_{[x,x+1)} H_p (t) \Pi^A_{[x',x'+1)} \right\| \le  \sum_{i=1}^N \left\| \Pi^A_{[x,x+1)} h_i(t) \Pi^A_{[x',x'+1)} \right\| .\nonumber
\end{eqnarray}
Because $A$ is given by $A=\sum_{i=1}^N a_i$ with $\|a_i\|=1$ for $i\in \Lambda$,  we have
\begin{eqnarray}
&\left\| \Pi^A_{\ge x} \Gamma^{(q)} \Pi^A_{\le x'} \right\| \le \|\Gamma^{(q)}\| \for |x'-x| \le 2 q,\nonumber\\
&\left\| \Pi^A_{\ge x} \Gamma^{(q)} \Pi^A_{\le x'} \right\| = 0 \for |x'-x| \ge 2 q\nonumber
\end{eqnarray}
for any $q$-local operator $\Gamma^{(q)}$.
We thereby obtain
\begin{eqnarray}
\fl \left\| \Pi^A_{[r_t x, r_t x+r_t)} h_i (t) \Pi^A_{[r_t x',r_t x'+r_t )} \right\| = \left\| \Pi^A_{[r_t x, r_t x+r_t)}( h_i (t)- \Gamma^{(q)} )\Pi^A_{[r_t x',r_t x'+r_t )} \right\|  
\end{eqnarray}
for $i\in \Lambda$ and $\forall \Gamma^{(q)}$ as long as $q\le r_t (|x'-x| -1)/2$.

Because $h_i$ is supported on the spin $i$,  
we can apply Theorem~\ref{Thm:exp_dynamics} for $q_0=1$, and hence 
\begin{eqnarray}
&\left\| \Pi^A_{[r_t x, r_t x+r_t)} h_i (t) \Pi^A_{[r_t x',r_t x'+r_t )} \right\| \nonumber\\
\le& \inf_{\Gamma^{(q)}\in \mathcal{U}(q), q\le r_t (|x'-x| -1)/2 } \| h_i (t)- \Gamma^{(q)} \| \nonumber\\
\le & 8 \|h_i \|  \lceil \kappa t \rceil 
\exp \biggl [ -\frac{1}{\xi} \Bigl ( \frac{1}{r_t}  \Bigl\lfloor \frac{r_t (|x'-x| -1)}{2} \Bigr\rfloor -1 \Bigr) \biggr]  \nonumber\\
=& 8\lceil \kappa t \rceil  \exp \left( - \frac{|x'-x|-5}{2\xi}  \right)  .
\end{eqnarray}
Therefore, we obtain
\begin{eqnarray}
\left\| \Pi^A_{[r_t x, r_t x+r_t)} H\Pi^A_{[r_t x',r_t x'+r_t )} \right\| &\le\sum_{i=1}^N \left\| \Pi^A_{[r_t x, r_t x+r_t)} h_i (t) \Pi^A_{[r_t x',r_t x'+r_t )} \right\| \nonumber\\
&\le  8N \lceil \kappa t \rceil  \exp \left( - \frac{|x'-x|-5}{2\xi}  \right)  ,
\end{eqnarray}
which finally yields the inequality~(\ref{t:eff}) with 
$
C_v =  8N e^{5/(2\xi )}\lceil \kappa t \rceil
$
and
$
\mu = 1/ (2\xi).
$

%%%%%%%%%%%%%%%%%%%%%%%%%%%%%%%%%%%%%%%%%%%%%%%%%%%%%%%%%%%%%%%%%%%%%%%%%%%%%%%%%%%%%%%%%%%%%%%%%%%%%%%%%%%%%%%%
%%%%%%%%%%%%%%%%%%%%%%%%%%%%%%%%%%%%%%%%%%%%%%%%%%%%%%%%%%%%%%%%%%%%%%%%%%%%%%%%%%%%%%%%%%%%%%%%%%%%%%%%%%%%%%%%
%%%%%%%%%%%%%%%%%%%%%%%%%%%%%%%%%%%%%%%%%%%%%%%%%%%%%%%%%%%%%%%%%%%%%%%%%%%%%%%%%%%%%%%%%%%%%%%%%%%%%%%%%%%%%%%%
%%%%%%%%%%%%%%%%%%%%%%%%%%%%%%%%%%%%%%%%%%%%%%%%%%%%%%%%%%%%%%%%%%%%%%%%%%%%%%%%%%%%%%%%%%%%%%%%%%%%%%%%%%%%%%%%
%%%%%%%%%%%%%%%%%%%%%%%%%%%%%%%%%%%%%%%%%%%%%%%%%%%%%%%%%%%%%%%%%%%%%%%%%%%%%%%%%%%%%%%%%%%%%%%%%%%%%%%%%%%%%%%%

%

%

%

%
%

%\appendix

\end{document}